\documentclass{aa}

\usepackage{graphics}

\begin{document}

   \thesaurus{12(02.07.1; 02.08.1; 03.13.1; 12.03.4; 12.12.1)}

   \title{Dynamical Scaling of Matter Density Correlations \\
          in the Universe.} 

   \subtitle{An Application of the Dynamical Renormalization Group.}

   \author{A. Dom\'\i nguez\inst{1}
           \and
           D. Hochberg\inst{1}
           \and
           J. M. Mart\'{\i}n--Garc\'{\i}a\inst{1}
           \and
           J. P\'erez--Mercader\inst{1} 
           \and
           L. S. Schulman\inst{2}
          }

   \offprints{D. Hochberg}

   \institute{Laboratorio de Astrof\'{\i}sica Espacial 
              y F\'{\i}sica Fundamental (INTA-CSIC)\\
              Apartado 50727, 28080 Madrid, Spain\\
          \and
              Department of Physics,\\ 
              Clarkson University, Potsdam, New York 13699-5820, USA
             }

   \date{Received Month day, year; accepted Month day, year}

   \maketitle

   \begin{abstract}

   We show how the interplay of non--linear dynamics, self--gravity, and
   fluctuations leads to self--affine behavior of matter density
   correlations quite generically, i.e., with a power law exponent
   whose value does not depend in a very direct way on the specific
   details of the dynamics.
   This we do by means of the
   Renormalization Group, a powerful analytical tool for extracting 
   asymptotic
   behavior of many-body systems.

     \keywords{ Gravitation --
                Hydrodynamics --
                Methods: analytical --
                Cosmology: theory --
                Large-scale structure of Universe
              }

   \end{abstract}

\section{Introduction.}

The Universe at short and moderate distance scales is inhomogeneous,
being filled by numerous structures. In fact the clumpy structure of matter
in the Universe extends for over 15 orders of magnitude in linear
size, from stars to the largest clusters of galaxies, and for about 18
orders of magnitude in mass (Ostriker \cite{Ostriker}).  At scales from galaxies
to clusters of galaxies and even superclusters, the Universe exhibits
self-similar, fractal behavior (Mandelbrot \cite{Mandelbrot}; 
Peebles \cite{Peebles}).  At the
largest scales, those probed by COBE observations (Smoot \cite{Smoot}), for
example, the Friedmann-Robertson-Walker (FRW) cosmology which is based
on the hypotheses of homogeneity and isotropy, provides adequate
description of many uncorrelated observations.  Isotropy is broken by
the primordial fluctuations in the matter density which subsequently
become amplified by the gravitational instabilities.

In this paper we explore the observed self-affinity and scaling in the
density correlation function and tackle them with the renormalization group, 
a tool originating in quantum field theory 
(Gell-Mann \& Low \cite{Gellmannlow}), then extended to 
condensed matter physics (Wilson \& Kogut \cite{Wilson}), 
later on  extended to problems
in hydrodynamics (Forster, Nelson and Stephen \cite{Forster}) 
and applied to surface growth (Kardar, Parisi and Zhang \cite{KPZ}) 
and in the last few years 
applied to gravitation and cosmology (P\'erez-Mercader et al. \cite{us}).

Our starting point will be the non-relativistic hydrodynamic equations
governing the dynamical evolution of an ideal self-gravitating
Newtonian fluid in a FRW background. Through a
series of steps involving physically justified approximations, these
equations reduce to a single evolution equation for the 
cosmic fluid's velocity
potential. At this point one has a single deterministic hydrodynamic
equation. To this we add a noise source which represents
the influence of fluctuations and dissipative processes 
on the evolution of the fluid,
arising from, but not limited to, viscosity, turbulence, explosions,
late Universe phase transitions, gravitational waves, etc.
The resulting equation is recognized as a cosmological
variant of the Kardar-Parisi-Zhang (KPZ) equation (Kardar et al. \cite{KPZ}) 
which is the simplest
archetype of nonlinear structure evolution and has been studied
extensively in recent years in the context of surface 
growth phenomena (Barab\`asi \& Stanley \cite{Barabasi}). 
While the essential steps involved in arriving
at this KPZ equation are reviewed in Section \ref{Hydro}, we refer the reader
to the concise details of its complete 
cosmological-hy\-dro\-dy\-na\-mi\-cal derivation in 
Buchert et al. \cite{DomBuchPM}.
The relevance of stochastic fluctuations in structure formation in the
Universe has been addressed in Berera \& Fang \cite{Berera}, 
where the r\^ole of the KPZ equation was emphasized.
For FRW cosmologies with curved spatial sections, this 
equation contains a time-dependent
mass-term. In Section \ref{Adiabatic} we discuss briefly the 
adiabatic approximation
which allows one to treat this mass term as a constant during most
of the expansion of the background cosmology. 

Having thus reduced the essential hydrodynamics to a
single dynamical stochastic equation, we turn to the results of a
dynamical renormalization group analysis to calculate the precise form
of the density-density correlation function using the KPZ equation.
To do so we proceed first by extracting the renormalization group
equations governing the evolution of couplings appearing in the
cosmological KPZ equation with respect to changes in scale and
integrating out the short distance degrees of freedom.  As we are
interested in the behavior of the system at ever larger scales, we
will run the renormalization group (RG) equations into the large
scale (infrared) limit, a process known as coarse-graining. The fixed
points of these RG equations determine the long--time, long--distance
behavior of the system and their subsequent analysis yields both the
power law form of the correlation functions as well as their explicit
exponents. In particular, the value of the galaxy-galaxy correlation
function exponent $\gamma \sim 1.8$ is calculated, and its value
understood in terms of spatio-temporal correlated noise. A summary of
our results and discussion are presented in the closing section.

\section{From Hydrodynamics to the KPZ Equation.}
\label{Hydro}

Structure formation in the Universe in the range of a few Mpc up to
several hundred Mpc can be modelled as the dynamical evolution of an
ideal self-gravitating fluid and studied within the framework of
General Relativity. However, we will restrict our treatment to the
case of Newtonian gravity for a number of physically justifiable
reasons. First and foremost, the length scales involved in large scale
structure formation by matter after decoupling from radiation are
smaller than the Hubble radius (of order $\approx 3000$Mpc at the
present epoch), so that general relativistic effects are negligible.
Secondly, it is known observationally that matter flow velocities are
much smaller than the speed of light and that this non-relativistic
matter plays a dominant r\^ole over radiation in the epoch of structure
formation. Thus, the evolution in space and time of large scale
structure in the Universe will be described adequately by the
Newtonian hydrodynamic equations for a fluid whose components are in
gravitational interaction in an expanding background metric
\footnote{The resolution length must be large enough so to as
  to legitimize the use of continuum equations (i.e., to validate the
  long-wavelength, or hydrodynamic limit) for the description of the
  matter density, peculiar velocity and acceleration fields. This
  requirement implicitly singles out a lower length scale which we may
  take to be greater than roughly the mean galaxy-galaxy separation
  length ($\sim 5$ Mpc).}. The connection with observations and the
subsequent phenomenology is made by writing these equations in terms
of the comoving coordinates $\vec x$ and coordinate
time $t$, and are (Peebles \cite{Peebles})

\begin{equation}\label{contin}
\frac{\partial \delta}{\partial t} + \frac{1}{a}{\vec \nabla}\cdot
[(1 + \delta){\vec v}] = 0,
\end{equation}

\begin{equation}\label{Euler}
\frac{\partial {\vec v}}{\partial t} + H{\vec v} + \frac{1}{a}
({\vec v}\cdot {\vec \nabla}){\vec v} = {\vec w} -\frac{1}{a\rho}
{\vec \nabla}p,
\end{equation}

\begin{equation}\label{Poisson}
{\vec \nabla}\cdot {\vec w} = -4\pi G a {\bar \rho(t)}\delta,\,\,\,
{\vec \nabla} \times {\vec w} = 0,
\end{equation}

\noindent
where $\vec{v}(\vec{x},t) $ and $\vec{w}(\vec{x},t) $ are the peculiar  
velocity and acceleration fields of the fluid, respectively.  
In the above equations, $H= \dot{a}(t)/a(t)$ is the Hubble  
parameter, $a(t)$ is the scale factor for the
underlying cosmological background and $\bar{\rho}(t)$ is the  
(time--dependent) average
cosmological density. Thus, the local density is  
$\rho = {\bar \rho}(1 + \delta)$ and the dimensionless density
contrast is $\delta = \frac{\rho - {\bar \rho}}{\bar \rho}$. 

In the linear regime and for dust ($p = 0$), the vorticity of the
velocity field $\vec v$ damps out rapidly by virtue of the background
expansion, and therefore ${\vec v}$ can be
derived from a velocity potential $\psi$, ${\vec v} = -\nabla \psi$, in
the long-time limit. Furthermore, parallelism between $\vec w$ and
$\vec v$ holds, i. e.,

\begin{equation}\label{Zeldo}
\vec{w} = F(t) \vec{v}
\end{equation}
where $F(t)$ is the long-time limit solution of the following
Ricatti equation

\begin{equation}\label{F}
\dot{F}= 4\pi G{\bar \rho(t)} - HF - F^2.
\end{equation}

Following the Zel'dovich approximation, we make the assumption that in
the weakly non-linear regime, the parallelism condition (\ref{Zeldo})
continues to hold, with $F(t)$ given as the solution of (\ref{F}) but
we now employ the fully non-linear Euler equation (\ref{Euler}) with
pressure term included. That is, we assume that the non-linearities
and pressure have not yet had enough time to destroy the
alignment in the acceleration and velocity fields of the cosmic
fluid. Assuming also that the pressure is a function of the density,
$p = p(\rho)$, one arrives at the following equation for the velocity
potential (Buchert et al. \cite{DomBuchPM})

\begin{equation}\label{KPZ}
\frac{\partial \psi}{ \partial t}= \nu f_1(t) \nabla^2 \psi +
\frac{\lambda}{2} f_2(t)  
(\nabla \psi)^2 +
\frac{f_3(t)}{T}\psi+\eta(\vec{x},t).
\end{equation}

\noindent
where we have also added a stochastic source or noise $\eta$ (see a
few lines below for a discussion). The three dimensionless functions
of time, $f_1$, $f_2$ and $f_3$ are

\begin{equation}\label{f1}
f_1(t)=\frac{p'(\bar \rho)\, F(t)}{4 \pi G \bar{\rho}(t) a^2(t)} \frac{1}{\nu},
\end{equation}

\begin{equation}\label{f2}
f_2(t)=\frac{1}{a(t)\lambda},
\end{equation}

\begin{equation}\label{f3}
f_3(t)= (F(t) - H(t))T.
\end{equation}

The positive constants $\nu,\lambda$ and $T$ are introduced to carry
dimensions. One can think of them as typical values of the
corresponding time-dependent coefficients during the epoch we are
interested in. The reason why we introduce them here explicitly
(instead of setting them equal to unity) will become clear in Section
\ref{RGEqs}.  The $f_1$-term arises from the pressure term in the Euler
equation which we have expanded to lowest order about the
zero-pressure limit. Note that higher-order terms in this Taylor
expansion will yield higher-derivative terms $(O(\nabla^2 \psi)^2)$
involving quadratic and higher powers of the field $\psi$. The
$f_2$-term is simply the convective term from the original Euler
equation (\ref{Euler}) written in the new variables, while the
$f_3$-term entails the competition between the damping of
perturbations due to the expansion $(-H\psi)$ and the enhancement due
to self-gravity $(+F\psi)$.

The Zel'dovich approximation (\ref{Zeldo}) together with 
the Poisson equation yield an important 
relationship between the density contrast and $\nabla^2 \psi$, 
the Laplacian of the velocity potential, namely

\begin{equation}\label{proportionality}
\delta(\vec{x},t) = \frac{F}{4\pi G a {\bar \rho}} \nabla^2 \psi(\vec{x},t),
\end{equation}

\noindent
which shows that the density contrast tracks the 
divergence of the peculiar velocity.
Because of this identity, we are able to calculate density correlation
functions in terms of (derivatives of) velocity potential correlations.
This is why it is worthwhile investigating the 
scaling properties of the KPZ equation in detail.

The noise $\eta(\vec{x},t)$ appearing in (\ref{KPZ}) represents the
effects of random forces acting on the fluid particles, including the
presence of dynamical friction, and also of degrees of freedom whose
size is smaller than the {\it coarse--graining length} (indeed, the
very fact that we are using continuous fields to describe the dynamics
of a system of discrete particles means that we are implicitly
introducing a coarse--graining length, such that the details below this
resolution length are not resolvable. (The relevance of this 
length-scale
will become clear in section \ref{RGEqs}.) There exist a number of physical
processes on various length and time scales that contribute to an
effective stochastic source in the Euler equation. Indeed, any
dissipative or frictional process leads to a stochastic force, by
virtue of the fluctuation-dissipation theorem. So, fluid viscosity and
turbulence should be accountable to some degree by adding a noise
term in the dynamical equations.  Early and late Universe phase
transitions, the formation of cosmic defects such as strings, domain
walls, textures, are sources for a noisy fluctuating background, as
are also the primordial gravitational waves and gravitational waves
produced during supernovae explosions and collapse of binary systems.
Another way to visualize the noise is to imagine a flow of water in
a river; there are stones and boulders in the river-bed which
obviously perturb the flow and have an impact on its nature. The noise
can be thought of as a means for modelling the distribution of
obstacles in the river bed. In cosmology the obstacles in the
river-bed represent, e.g., density fluctuations and the fluid is the
matter flowing in this ``river''.  The noise is phenomenologically
characterized by its average value and two--point correlation function
as follows,

\begin{equation}
\left\langle \eta(\vec{x},t) \right\rangle =0
\end{equation}

\begin{equation}
\left\langle   \eta(\vec{x},t)    \eta(\vec{x}',t')       
\right\rangle =2 D(\vec{x},\vec{x}';t,t'),
\end{equation}

\noindent
with $D(\vec{x},\vec{x}';t,t')$ a given function of its arguments (see  
below). All higher cumulants vanish. The noise is thus Gaussian. However,
the velocity field need not be (and in general, will not be)
Gaussian as a consequence of coarse-graining of the dynamics.

The stochastic hydrodynamic 
equation (\ref{KPZ}) can be further cast into
\begin{equation}\label{cosmoKPZ}
\frac{\partial \Psi}{ \partial \tau}= \nu \nabla^2 \Psi +  
\frac{1}{2}\lambda (\nabla \Psi)^2 -
m^2(\tau) \Psi+\tilde{\eta}(\vec{x},\tau),
\end{equation}

\noindent
by means of the following simple change of variables and redefinition of  
the ``physical--time'' $t$ into a ``conformal--time'' $\tau$

\begin{equation}\label{functions}
\Psi(\vec{x},\tau) \equiv \frac{f_2(t(\tau))}{f_1(t(\tau))}
\psi(\vec{x},t(\tau)),
\end{equation}

\begin{equation}\label{conformalTime}
\tau=\int_{t_0}^{t} d t' f_1(t').
\end{equation}

\noindent
The linear (mass) term in equation (\ref{cosmoKPZ}) is given by

\begin{equation}\label{mass}
m^2(\tau) = -\frac{d}{d\tau} \ln \left[ \frac{f_2(\tau)}{f_1(\tau)} \exp 
\left( {1 \over T} \int_{t_0(\tau_0)}^{t(\tau)}d t'\  f_3(t') \right)  \right],
\end{equation}

\noindent
and the noise transforms into 
\begin{equation}
\tilde{\eta}(\vec{x},t) = \frac{f_2(t)}{f_1(t)^2}\eta(\vec{x},t).
\end{equation}

Finally, Eq. (\ref{cosmoKPZ}) is a generalization to a cosmological  
setting of the Kardar--Parisi--Zhang (KPZ) equation for surface  
growth (Kardar et al. \cite{KPZ}), 
and differs from the standard KPZ equation in the linear  
(mass) term which, as seen from Eq. (\ref{mass}), originates in the  
expansion present in the background cosmology.  
Here, $\nu$ plays the r\^ole of a diffusion constant while $\lambda$
is proportional to the average ``speed of growth'' of $\psi$.

\section{Adiabatic Approximation.}
\label{Adiabatic}

In this section we briefly comment on some results about the
time--dependent mass term (\ref{mass}) of equation (\ref{cosmoKPZ}).
The detailed analysis that justifies our assertions is carried out in
Buchert et al. \cite{DomBuchPM}. 
 
For equations of state of the form $p = \kappa \rho$, the
time--dependent mass term (\ref{mass}) is identically zero for flat
FRW backgrounds, but evolves with time in closed and open FRW
backgrounds, keeping a constant sign during the epoch of interest:
$m^2 > 0$ in open backgrounds and $m^2 < 0$ when the background is
closed.

There are two time scales in our problem: an intrinsic time scale
associated with the dynamical evolution prescribed by equation
(\ref{cosmoKPZ}), which is determined by the values of the
coefficients in it, and an extrinsic time scale associated with the
background expansion, which is determined by the parameters defining
the background cosmology. A numerical study of the function $m^2(t)$
for the physically interesting range of parameters, reveals that its
relative variation over these two time scales is small, which
justifies an ``adiabatic'' approximation: we can assume $m^2(t)$ to be
a constant, rather than a function of time, when performing the
Renormalization Group analysis in the next section.  This assumption
greatly simplifies the analysis, and the presence of a non-zero mass
term in the KPZ equation results in a much richer renormalization
group trajectory flow and fixed point structure than that
corresponding to the massless case, as we will see below.

\section{The Renormalization Group Equations}
\label{RGEqs}

The starting point for the subsequent analysis is the KPZ equation for
the velocity potential, eq. (\ref{cosmoKPZ}). Because this is a 
nonlinear equation, and
because there are fluctuations, as represented by the stochastic noise
term, it is to be expected on general grounds that renormalization
effects will modify the coefficients appearing in this equation.  Due
to the existence of fluctuations, the coefficients in equation
(\ref{cosmoKPZ}) $(\nu,\lambda,m^2)$ do not remain constant with
scale: as is well known (Ma \cite{Ma}; Amit \cite{Amit}; Binney \cite{Binney})
there are
renormalization effects that take place and modify these coefficients
(or ``coupling'' constants); the modifications are calculable and can
be computed using dynamical renormalization group (DynRG) techniques.
The renormalization group is a standard tool developed for revealing
in a systematic way, how couplings change with scale in any physical
system under the action of (coarse)-graining. Moreover, it predicts
that all correlation functions go as power laws when the system is
near a fixed, or critical point, i.e., they exhibit self--similar
behavior.  However, the value of the scaling exponent changes from one
fixed point to another. {\it In fact, the values for the couplings at some
reference scale establish the fixed point to which the system will be
attracted to or repelled from, because these reference values will
belong to a specific ``basin of attraction".} We will look for the IR
stable fixed points since these reflect the system behavior
characteristic of long times and large distances and ceases to change
as we look at the system at ever larger scales.

Indeed, even before fluctuations are accounted for, one can 
easily obtain the so-called canonical scaling laws for the parameters
appearing in (\ref{cosmoKPZ}); these follow by
performing the simultaneous scaling transformation 
${\bf x} \rightarrow s {\bf x}$, $\tau \rightarrow s^z \tau$ and
$\psi \rightarrow s^{\chi} \psi$ and requiring the resultant equation to be 
{\it form-invariant}, that is, that the KPZ equation transforms to
a KPZ equation at the new, larger ($s > 1$) length and time scale. 
This simple requirement leads to scale dependence in the parameters
appearing in the original equation of motion as well as in the
correlation functions built up from the KPZ field $\psi(\vec{x},t)$.
When fluctuations and noise are ``turned on'',  renormalization  
effects change the scaling behavior of the couplings and 
correlation functions  
away from their canonical forms, in a way which depends on the  
basins of attraction for the fixed points of the DynRG equations of  
the couplings.

\subsection{Calculation of the RGEs.}

The RG transformation consists of an averaging over modes with momenta
$k$ in the range $\Lambda/s \leq k \leq \Lambda$ where $s > 1$ is the scale
factor for the transformation, followed by a dilatation of the length scale
${\bf x} \rightarrow s{\bf x}$ in order to bring the system back to its 
original size\footnote{The term ``renormalization group'' is actually a
misnomer, since due to the process of averaging over small distances and the
subsequent loss of information, the RG transformations do not have an inverse.
Consequently the technically correct name would have to be the
``renormalization semi-group''.}. 
Here, $\Lambda$ plays the r\^ole of an UV-momentum, or short distance, cut-off
characterizing the smallest  resolvable detail whose physics is to
be described by the dynamical equations. 
In classical hydrodynamics, this scale would be identified with the
scale in the fluid at and below which the molecular granularity of the medium
becomes manifest and the hydrodynamic limit breaks down.
The renormalization group 
transformation has the important property that it becomes 
an {\em exact} symmetry
of the physical system under study whenever that system approaches or is near
a critical point, because it is near the critical point (or points) where
the system exhibits scale-invariance or, equivalently, where the
system can be described in terms of a conformal field theory. 
The hallmark for a system near criticality is that its
correlation functions display power law behavior. 
This means that the statistical properties of the 
system remain the same, except possibly up to a global dilatation or change of unit of length. We can
calculate the power law exponents by requiring that the dynamical 
equations remain invariant under the above RG transformation and under the
further change of scale

\begin{equation}\label{scaling}
x \rightarrow s x \,\,\,\, \,\,\,\, t \rightarrow s^z t,
\,\,\,{\rm and} \,\,\,\, \Psi \rightarrow s^{\chi} \Psi,
\end{equation}
where $z$ and $\chi$ are numbers which account for the 
response to the re-scaling.
By eliminating $s$ from the above, we arrive at the
fact that the two-point correlation function for the velocity potential
scales as

\begin{equation}\label{corscale}
\langle  \Psi (\vec{x},t)    \Psi (\vec{x}',t')  \rangle \propto  
\mid \vec{x} - \vec{x}'\mid^{2 \chi}
f\left( \frac{ \mid t - t'\mid}{ \mid \vec{x} - \vec{x}'\mid^z} \right),
\end{equation}

\noindent
where $\chi$ is the roughness exponent, $z$ the dynamical exponent,  
and the scaling function $f(u)$ has the following asymptotic  
behavior (see, e. g., Barab\`asi \& Stanley \cite{Barabasi}):

\begin{equation}\label{limf1}
\lim_{u\rightarrow \infty} f(u) \rightarrow u^{2 \chi /z},
\end{equation}

\begin{equation}\label{limf2}
\lim_{u\rightarrow 0} f(u) \rightarrow {\rm constant}.
\end{equation}

Notice that because of eq. (\ref{scaling}), large $s$ means 
going to the large distance or infrared (IR) and (for $z>0$) long time limit,  
while small $s$ corresponds to the short distance or ultraviolet (UV) and
short time limit.

Making use of the relations Eq.(\ref{functions}) and Eq.  
(\ref{proportionality}), one can immediately obtain the scaling  
behavior of the 2--point correlation function for the density  
contrast, and thus study the asymptotic behavior of this function in  
different regimes. In the following we will implement this
procedure. First, we obtain the scaling or RG equations for the
couplings by imposing form-invariance on the equation (\ref{KPZ}), then
by using the property of constancy of the couplings near fixed points, we
obtain and calculate the fixed points themselves and the corresponding
values of the exponents.

We now characterize the Gaussian noise, $\eta(\vec{x}, t)$. This is
done by choosing the noise correlation function. Here we will use
colored or correlated noise, whose Fourier transform satisfies

\begin{eqnarray}
\langle \tilde{\eta}(\vec{k},\omega) \rangle & =& 0, \nonumber \\
\langle \tilde{\eta}(\vec{k},\omega)\tilde{\eta}(\vec{k}^\prime,\omega^\prime)
\rangle  
& =&  
2\tilde{D}(k,\omega)(2\pi)^{d+1}\delta(k+k^\prime)\delta(\omega+\omega^\prime),
\nonumber \\
\tilde{D}(k,\omega) & = &  D_0+D_{\theta} k^{- 2 \rho} \omega^{- 2 \theta},
\label{noiseFourier}
\end{eqnarray}
where $D_0$ and $D_\theta$ are two couplings  
describing the noise strength, and the $\rho$, $\theta$ 
exponents characterize the  
noise power spectrum in the momentum and frequency domains. White  
noise corresponds to $D_\theta=0$.
The explicit functional form of the noise 
amplitude $\tilde D(k,\omega)$ reflects the fact that
correlated noise has power law singularities of the 
form written above (Medina el al. \cite{MHKZ}).

The solution of (\ref{cosmoKPZ}) can be carried out in Fourier space
where iterative and diagrammatic techniques may be developed 
(Medina et al. \cite{MHKZ}).  A standard  
perturbative expansion of the solution to Eq. (\ref{cosmoKPZ})  
coupled with the requirement of form--invariance and the property of  
renormalizability, lead to the following RG equations for the  
couplings: 

\begin{eqnarray}\label{RGEs}
\frac{{\rm d}m^2}{{\rm d}\log s} &=& zm^2, 
\nonumber\\
\frac{{\rm d}\nu}{{\rm d} \log s} &=& 
\nu[z-2-\frac{\lambda^2}{\nu^3}\frac{K_d}{4d}\Lambda^{d-2} V^{-2}\{(d-2V^{-1})D_0 \nonumber \\
&& +(d-2V^{-1}-2\rho)D_\theta V^{-2\theta}\sec(\theta\pi)(1+2\theta)\}], 
\nonumber \\
\frac{{\rm d} D_0}{{\rm d}\log  
s}&=&D_0(z-2\chi-d)+\frac{\lambda^2}{\nu^3}\frac{K_d}{4}
\Lambda^{d-2}V^{-3}\times \\
& & [D_0^2+2D_0D_\theta V^{-2\theta}\sec(\theta\pi)(1+2\theta) \nonumber \\
&& +D_\theta^2 V^{-4\theta}\sec(2\theta\pi)(1+4\theta)], \nonumber \\
\frac{{\rm d} D_\theta}{{\rm d}\log  
s}&=&D_\theta[z(1+2\theta)-2\chi-d+2\rho], \nonumber \\
\frac{{\rm d}\lambda}{{\rm d}\log  
s}&=&\lambda[\chi+z-2 \nonumber \\
&& -\frac{\lambda^2D_\theta}{\nu^3}
\frac{K_d}{d}\Lambda^{d-2}V^{-3-2\theta}\theta(1+2\theta)\sec(\pi\theta)]. 
\nonumber
\end{eqnarray}

\noindent
The full details of the calculation of these equations will be
presented elsewhere (Mart\'\i n--Garc\'\i a \& P\'erez-Mercader \cite{jmm}). 
The calculation of the RGE's for the {\it massless} KPZ equation 
in the presence of colored noise is given in (Medina et al. \cite{MHKZ}).
Here, $s$ is the scale factor of Eq. (\ref{scaling}), 
and $K_d = \frac{S_d}{(2\pi)^d}$ is a dimensionless
geometric factor proportional to the surface area of the $d$-sphere $S_d$,
where $d$ is the spatial dimension.  This  
set of equations describes how the coupling constants 
$\nu,\lambda,m^2,D_0,D_\theta$ evolve
as one varies the scale at which the system is studied, a process  
commonly referred to in the condensed matter literature as ``graining".   
In the present context we will be interested in the coarse-graining
behavior, since we seek to uncover the behavior of 
our cosmological fluid as 
one goes to larger and larger scales.
In actual practice we will investigate the coarse-graining flow in a
two-dimensional (white noise) or three-dimensional (correlated noise)
parameter space. This is because one may cast the above set of RG equations
in terms of a smaller, yet equivalent set, by employing the 
dimensionless couplings defined as
$V=1+\frac{m^2}{\nu \Lambda^2}$ ,
$U_0=\lambda^2D_0K_d\Lambda^{d-2}/\nu^3$, and
$U_\theta=\lambda^2 D_\theta K_d \Lambda^{d-2-2\rho-4\theta}/\nu^{3+2\rho}$.
Doing so, one arrives at the following reduced set of
renormalization group equations:
\begin{eqnarray} \label{reduced}
\frac{{\rm d}U_0}{{\rm d}\log s}&=&
  (2-d)U_0 + \frac{U_0^2}{4dV^3}\left[d+3(dV-2)\right] \nonumber \\
&& + \frac{U_\theta^2}{4V^{3+4\theta}}\sec(2\pi\theta)(1+4\theta)
\nonumber \\
&& +\frac{U_0U_\theta}{4dV^{3+2\theta}}(1+2\theta)\sec(\pi\theta) \times
\nonumber \\
&& \qquad [2d-8\theta+3((d-2\rho)V-2)]
\\
\frac{{\rm d}U_\theta}{{\rm d}\log s}&=&
  (2-d+2\rho+4\theta)U_\theta+
  \frac{U_0U_\theta}{4dV^3}(dV-2)(3+2\theta)
\nonumber \\
&&
 +\frac{U_\theta^2}{4dV^{3+2\theta}}(1+2\theta)\sec(\pi\theta) \times
\nonumber \\
&& \qquad [-8\theta+((d-2\rho)V-2)(3+2\theta)] 
\nonumber
\\
\frac{{\rm d}V}{{\rm d}\log s}&=&(V-1)
  [2+\frac{1}{4dV^3}
       [(dV-2)U_0 \nonumber \\
&&   +((d-2\rho)V-2)V^{-2\theta}(1+2\theta)\sec(\pi\theta) U_\theta ]
  ]
\nonumber
\end{eqnarray}

The dimensionless couplings $U_0,U_{\theta}$ measure the strength of
the ``roughening'' effect due to the combined action of noise and
the non-linearity against the ``smoothing'' tendency of 
the diffusive term in equation (\ref{cosmoKPZ}):
these couplings grow when either $\lambda$ or the noise intensity
(as given by the parameters $D_0$ and $D_\theta$) grow, and they become
smaller when the diffusive action (measured by $\nu$) grows.
The dimensionless coupling $V$ measures the competition between the
diffusion and mass terms. Indeed, neglecting for the moment the
noise and nonlinear terms, one can write (\ref{cosmoKPZ}) 
in Fourier space as
\begin{equation}
\frac{\partial {\tilde \Psi}}{\partial \tau}
              = -(\nu k^2 + m^2){\tilde \Psi}.
\end{equation}
Now, if $m^2 \geq 0$ (i.e., $V \geq 1$), the perturbations in the field
$\Psi$ are damped. But if on the other hand, $m^2 < 0$ 
($V < 1$), there exists a length scale $L = \sqrt{-\frac{\nu}{m^2}}$
below which the perturbations are damped, but above which they grow.
The case $V < 0$ corresponds to a length scale $L < {\Lambda}^{-1}$: this
scale is {\em smaller} 
than the minimum resolvable length scale in the problem at hand, 
and therefore the
diffusive term becomes unimportant. We remark in passing that the limiting
behavior in the RG flow as $V \rightarrow 0^{+}$ leads to a technical
requirement: one must assume that the diffusive term is non-negligible
(i.e., $\nu \neq 0$) in order to compute the RG equations.

From the above equations one calculates 
their fixed points to which the  
graining flow drives the couplings. The corresponding fixed point 
exponents $z$ and $\chi$
follow from substituting the fixed point solutions so
obtained
back into the previous set of RG equations (\ref{RGEs}).    
These exponents control and determine the 
{\it calculated}  
asymptotic behavior of the correlation functions.
The system will be attracted to fixed points in the
IR or UV regimes depending on whether they are IR-attractive
or UV-attractive although, as with any 
autonomous set of differential equations, other possibilities
exist.

\section{Fixed Points.}

The fixed points, collectively denoted by $g^*$, are the solutions to
the system of algebraic equations obtained by setting the right hand
sides of equations (\ref{reduced}) equal to zero.  They control the
stability properties of the system of DynRG equations and therefore
they characterize the asymptotic behavior of correlation functions.
When the couplings attain their critical values, the system becomes
critical and the correlation functions enter into power law regimes,
characteristic of physical systems in the critical state where they
are scale invariant.  {\it Which fixed point the system is attracted to or
repelled from depends on the basin of attraction where the initial
conditions for the DynRG equations are located.}  As the couplings
evolve under graining into their critical values, the $n$--point
correlation functions approach power law (or scaling) forms because
they are the solution to first order quasi--linear partial
differential equations (the Callan-Symanzik (CS) equations
(Amit \cite{Amit}; Binney \cite{Binney})) whose characteristics are the 
solutions to Eqns.
(\ref{RGEs}).  This CS equation simply expresses the fact that the
correlation functions are invariant under the renormalization group
transformation. In other words, the system is scale-invariant at its
fixed points, hence, correlation functions must take the form of
algebraic power laws, since these are the only mathematical functions
that are scale-invariant.  The $physics$ of each of these fixed points
is different, because it depends to a large extent on the magnitude
and sign of the scaling exponent as well as on the attractive or
repulsive nature of the fixed point.

\begin{figure}
   \resizebox{\hsize}{!}{\includegraphics{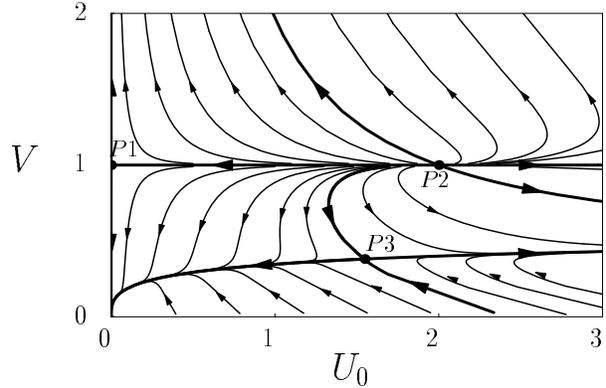}}
   \caption{Flow diagram in ($U_0$,$V$) space. Arrows indicate IR
            ($s\rightarrow \infty$) flow.}
   \label{figure1}
\end{figure}

\begin{figure}
   \resizebox{\hsize}{!}{\includegraphics{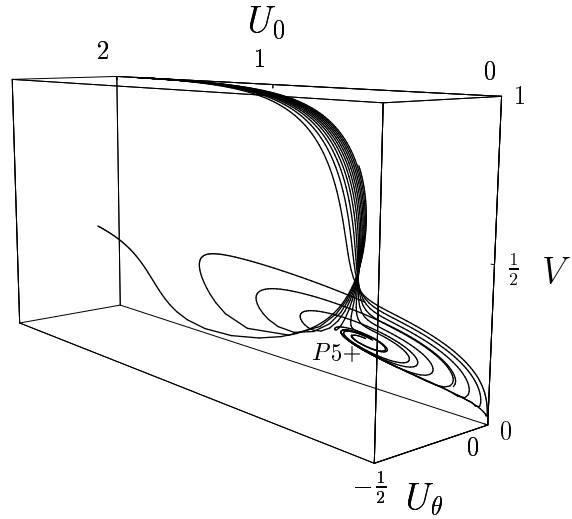}}
   \caption{Flow diagram in ($U_\theta$, $U_0$, $V$) space around P5+.}
   \label{figure2}
\end{figure}

The flows and the fixed points of our set of RG equations
are represented in Fig.~\ref{figure1}, for  
white noise and Fig.~\ref{figure2} for colored noise, while the features of our
fixed point analysis are conveniently summarized in the adjoining Table 1.

\begin{table*}
\caption{Characteristics of the fixed points.}
\begin{tabular}{|c|c|c|c|c|c|}

\hline Point & Position $(U_\theta^*,U_0^*,V^*)$ & $(z,\chi)$ & $\gamma=4-2\chi$ 
& IR-eigenvalues & Class \\

\hline P1 & $(0,0,1)$ & $(2,-1/2)$ & $5$ & $(4.7,-1,2)$ & Saddle point \\ 

\hline P2 & $(0,2,1)$ & $(13/6,-1/6)$ & $13/3$ & $(5.2,1,2.2)$ & IR-repulsive \\

\hline P3 & $(0,1.55,0.38)$ & $(0,2)$ & $0$ & $(-1.7,2.0,-15.1)$ & Saddle point \\

\hline P4$\pm$ & \parbox{5.5cm}{$V^*=1, \ (U_0^*,U_\theta^*)$ depend on $(\rho,\theta)$ Some pairs are not allowed by a discriminant condition$^*$} & \parbox{2.75cm}{Depend on $(\rho,\theta)$ $\chi\le 0$ or $\chi\ge 4/3$} & \parbox{2.5cm}{Depends on $(\rho,\theta)$ $\gamma\ge 4$ or $\gamma\le 4/3$} & \parbox{2.5cm}{Sign depends on $(\rho,\theta)$}& \\[4mm]

\cline{2-6} & For $\rho=2.65, \ \theta=0.1$ & & & & \\
& P4+ \ \ \ \ \ \ \ $(4.2,6.4,1)$ & $(0.64,1.53)$ & $0.94$ & $(-6.2,1.8,0.64)$ & Saddle point \\
& P4-- \ \ \ \ \ \ \ \ $(3.4,1.7,1)$ & $(0.62,1.52)$ & $0.96$ & $(-5.5,-1.6,0.62)$ & Saddle point \\

\hline P5$\pm$ & \parbox{5.5cm}{$(U_\theta^*,U_0^*,V^*)$ depend on $(\rho,\theta)$ Some pairs are not allowed by a discriminant condition$^*$. Essentially the allowed region is $\rho\in[0,7/2], \\ \theta\in[0,1/4)$} & \parbox{2.75cm}{$(0,\rho-3/2)\Rightarrow \chi\in[-3/2,2]$ which includes the range $(1.1,1.25)$ for $\chi$} & $7-2\rho$ & \parbox{2.5cm}{Sign depends on $(\rho,\theta)$}& \\

\cline{2-6} & For $\rho=2.65, \ \theta=0.1$ & & & & \\
& P5+ $(-0.1,0.43,0.19)$ & $(0,1.15)$ & $1.7$ & $(-0.4\pm 3.5i,-27.3)$ & IR-attractive \\
& P5-- $(3.0,-1.96,-0.55)$ & $(0,1.15)$ & $1.7$ & $(-19.2,8.5,-1.5)$ & Saddle point \\
\hline
\end{tabular}

\parbox{18cm}{$^*$ These fixed points are obtained as real solutions of a second degree 
algebraic equation, and therefore the set of pairs that give real fixed points
are restricted by a discriminant inequality.}

Remarks:
\begin{itemize}
\item Only P5+ has a basin of attraction, which is approximately
in the region $V\in(0,1),U_0\in(0, 1.5),U_\theta\in(-0.2,0)$.  
P2 has a large ``basin of repulsion'', and the other points are saddle points.
\item The trajectories outside the P5+ basin of attraction reach infinity or fall
to the singular $V=0$ plane at a finite ``time''.
\item Only with P5+ can we adjust the observed values $\gamma\in(1.5,1.8)$, 
because P5-- has $V^*<0$ in this region, which implies smaller length
scales than the one corresponding to the UV cut-off $\Lambda$.
\end{itemize}
\end{table*}

\subsection{The value of the Critical Exponents at each Critical  
Point. White Noise.}

In the case of white noise ($D_\theta=0$) we solve for the fixed points
of the (reduced set) of RG equations. Since $U^{*}_\theta = 0$ in this case, 
these fixed points will all lie
in a two-dimensional coupling space spanned by $(U_0,V)$.  
There are three fixed points, labelled
as $P1,P2$ and $P3$, whose coordinates are listed in Table 1. 
Two of them, $P1,P2$ lie in the line $V^* = 1$, 
corresponding to ${m^*}^2 = 0$ in the RG-evolved
KPZ equation, while $P3$ lies in the region bounded
between the $V = 1$ and $V = 0$ lines.
By substituting
these fixed point values  back into the original 
set of RG equations (\ref{RGEs}), we solve for
the corresponding fixed point exponents $\chi$ and $z$. 
These values are listed in the third column of Table 1.
Linearizing the RG 
equations about each one of their fixed points allows one to calculate
the infrared (IR) stability properties of the fixed points and thus 
characterize their behavior with respect to the coarse-graining. This
entails expanding the RG equations in 
$U_0 \rightarrow U^*_0 + \delta U_0$
and  $V_0 \rightarrow V^*_0 + \delta V_0$ to first order in the perturbations
and solving for $\delta U_0$ and $\delta V_0$. 
In general, the eigenvectors of this
linearized system will involve linear 
combinations of  $\delta U_0$ and $\delta V_0$.
The associated eigenvalues, which determine the stability properties of the
fixed point, are also listed in Table 1. A positive eigenvalue indicates
that the coarse-graining induces a flow away from the point along the
eigen-direction, while a negative eigenvalue indicates the point is
stable in the infrared, since the flow will be into the fixed point. 
The type, or class, of fixed point, whether it be
IR-attractive (all eigenvalues negative), IR-repulsive 
(all eigenvalues positive)
or a saddle point (mixed sign eigenvalues) is listed in Table 1.
The coarse-graining flow of the couplings in the white noise case
is given in Fig.~\ref{figure1} . These flow lines were calculated by integrating
numerically the set of two coupled first order differential equations
for $U_0$ and $V$, using general choices for the RG equation initial
conditions. 

\subsection{The value of the Critical Exponents at each Critical  
Point. Colored Noise.}

For colored or correlated noise ($D_\theta \neq 0$), we now deal with
a three dimensional space of dimensionless couplings $U_\theta, U_0$
and $V$.  Using the same procedure as discussed above, we solve for the
RG equation fixed points which lead to a total of seven fixed points,
including the same three points $P1,P2$ and $P3$ obtained in the white
noise limit.  Thus, allowing for colored noise yields four additional
fixed points which we will denote by $P4\pm$ and $P5\pm$, since they
arise in pairs.  Their coordinates are listed in the Table, together
with their
associated exponents $\chi$ and $z$ and IR eigenvalues and
stability properties under coarse-graining. Both, the positions of
these points and values of their exponents, depend in general upon the
values of the noise exponents $\rho$ and $\theta$, which parametrize
spatial and temporal correlations in the noise.  The pair of fixed
points labelled as $P4 \pm$ always lie in the plane $V^* = 1$, (i.e.,
${m^*}^2 = 0$) but their location within this plane varies with the
noise exponents. Moreover, any RG flow which starts off in this plane
will always remain in this plane (this plane acts as a separatrix).
It is important to point out that there are values of the noise
exponents $\rho,\theta$ which lead to {\it complex} values of the
fixed point coordinates for $P4\pm$.  This is because these fixed
points arise as solutions of a quadratic algebraic equation whose
discriminant can become negative for values of $\rho$ and $\theta$ in
certain domains in parameter space.  We must rule out such values of
the noise exponents because of physical reasons.  For the allowed values of
$\rho$ and $\theta$ (i.e., those that lead to real fixed points), we
find that $P4 \pm$ have exponents $z$ and $\chi$ that are complicated
functions of these parameters, but we have checked that for all allowed
values, either $\chi \leq 0$ or $\chi \geq \frac{4}{3}$.  The IR
eigenvalues and the nature of these fixed points depend on $\rho, \theta$.
For the choice shown, i.e., $\rho = 2.65$ and
$\theta = 0.1$, both $P4\pm$ are saddle points.  In fact, we have
confirmed that the pair $P4\pm$ will always be saddle points whenever
$\rho \in (2.60,2.75)$ and $\theta \in (0, .23)$. The reason for
choosing these particular exponent intervals will become clear in
Section~\ref{Scalingsection} .

The next pair of fixed points $P5\pm$ also have coordinates and exponents
depending on the noise exponents. As in the case of $P4\pm$, there are
values of $\rho,\theta$  (same as for $P4\pm$)
leading to complex couplings, which we rule out. 
The allowed values of the noise exponents
lie roughly in the respective intervals $\rho \in [ 0,\frac{7}{2} ]$ and
$\theta \in [ 0, \frac{1}{4} )$.
The exponents for $P5\pm$ are given by the  simple functions
$z = 0$ and $\chi = \rho -\frac{3}{2}$, and the stability
properties do not depend on the particular values of
the noise exponents within the above mentioned intervals. 
For the choice $\rho = 2.65$ and $\theta 
= 0.1$, $P5+$ is IR attractive while $P5-$ is an unstable saddle point.
We must nonetheless exclude $P5-$ from our consideration since it lies
below the plane $V = 0$, i.e., it corresponds to a renormalized
KPZ mass ${m^*}^2 < - \nu \Lambda^2$, which therefore exceeds the 
scale of the momentum cut-off
imposed on our perturbative calculation.

\section{Scaling properties of matter density correlations.}
\label{Scalingsection}

We can now begin to put together the results we have obtained in  
relation with the scaling properties {\it predicted} by the DynRG. Using  
equations (\ref{proportionality}) and (\ref{functions}) 
we can write for the 2--point correlation function of the density contrast
in comoving coordinates that

\begin{eqnarray}\label{xi}
\xi(|\vec{x}-\vec{y}|, t) 
&\equiv&
\left\langle \delta (\vec{x},t) \, \delta (\vec{y},t) \right\rangle \\
&=& \left(\frac{F}{4\pi Ga{\bar \rho}} \right)^2  \frac{f^2_1}{f^2_2}
\nabla_x^2 \nabla_y^2
\left\langle \Psi (\vec{x},t)  \Psi (\vec{y},t') \right\rangle \mid_{t'=t} \, .
\nonumber
\end{eqnarray}

\noindent
We see that the scaling behavior of $\xi(r)$ is determined  
once we know the scaling behavior for the equal time two--point  
correlation function of the density contrast for the velocity 
potential $\Psi (\vec{x},t)$,

\begin{equation}\label{cder}
C(|\vec{x}-\vec{y}|) \equiv
\left\langle \Psi (\vec{x},t)  \Psi (\vec{y},t) \right\rangle \, .
\end{equation}

\noindent
>From Eqns. (\ref{corscale}) and (\ref{limf2}), we have

\begin{equation}
\left\langle \Psi (\vec{x},t)  \Psi (\vec{y},t) \right\rangle \sim  
|\vec{x}-\vec{y}|^{2 \chi} \, .
\end{equation}

\noindent
Introducing (for convenience and notational consistency with the  
literature in critical phenomena) $\eta=2-d-2\chi$, we have

\begin{equation}
C(r) \sim r^{-(d-2+\eta)} \,\,;\,\, \tilde{C}(k) \sim k^{\eta -2}
\end{equation}

\noindent
which on using equation (\ref{xi}) above for equal times gives

\begin{equation}\label{corrFunct}
\xi(r) \sim r^{-(d+2+\eta)},
\end{equation}

\noindent
for the correlation function, and

\begin{equation}\label{powerSpec}
P(k) \sim k^{2+\eta},
\end{equation}

\noindent
for its Fourier transform, the power spectrum.

Equations (\ref{corrFunct}) and (\ref{powerSpec}) are now suitable  
for comparison with observations\footnote{Observation has shown that 
large scale structure depends on cosmological parameters such as
the age of the Universe $t_0$ and initial density $\Omega_0$. While these
parameters (and others) do not appear explicitly in our
calculations, they are there implicitely in quantities such
as $r_0$, the scale factor, whose calculation cannot
be carried out without a deeper knowledge of the noise
sources. Nevertheless, the renormalization group gives one
a way to calculate the power law scaling law and its exponent
as a function of which particular basin of attraction the
systems happens to start out in. This explains exactly what
we mean when we claim that no fine tuning of parameters is required
in order to account for the value for the scaling exponent
inferred from the observations. Irrespective of
where the system starts out in
a particular basin of attraction, in time it inevitably flows into that
basin's fixed point: this is very much a kind of independence
of initial conditions.}, since 
they are written in comoving  
coordinates, and therefore correspond to quantities directly  
inferred from observations and provided we assume that light
traces mass. Otherwise we would need to correct for bias, a strategy
we leave for a future publication.
From (\ref{corrFunct}) and the definition of the exponent $\eta$, we
see that the exponent $\gamma$ measured from large scale galaxy surveys,
where $\xi_{OBS}(r)\sim r^{-\gamma}$,
is calculable in term of the roughening exponent $\chi$ and is 
given by \footnote{Notice that when written in this way, the value of 
$\gamma$ seems to be independent of the spatial dimensionality
$d$. But this is fallacious, since $\chi$ itself depends on $d$ as
is seen from the RGE's.} 

\begin{equation}\label{gamma}
\gamma = 4 - 2\chi (\rho,\theta).
\end{equation}

Using our fixed point analysis we have computed all the exponents
$\chi$ and $z$ for all the fixed points (fourth column in Table 1). The predicted
values for $\gamma$ are tabulated in the Table as shown. Thus, at each
fixed point, we derive a power law for the density correlation function
$\xi(r) \sim r^{-\gamma}$. In the case of white noise alone, we see that
none of the three corresponding 
fixed points are capable of reproducing any of the inferred
values of $\gamma$
($1.5 \stackrel{<}{\sim} \gamma \stackrel{<}{\sim} 1.8$) from 
observations.
It is of interest to point out, however, that
the point $P3$ yields the exponent $\gamma = 0$, which means that the
correlation function is strictly constant at this point, i.e., the
matter distribution is perfectly homogeneous. However, $P3$ is
an IR unstable saddle point and a fine-tuning in the initial values
of $U_0$ and $V$
would be required in order to
have the system flow into it under coarse-graining.

For colored noise, the RG behavior becomes significantly richer since
we now must map out RG equation flow trajectories in a three
dimensional space of couplings (see Fig.~\ref{figure2}).  A glance at the
tabulated calculated values of the exponent $\gamma$ shows that $P5+$
is the only fixed point capable of yielding values of $\gamma$ within
the currently accepted range of values inferred from observations, by choosing $\rho \in
(2.60, 2.75)$ and any $\theta \in (0, .23)$, since $\chi = \chi(\rho)
= 7 -2\rho$ for $P5\pm$.  Although $\gamma$ is independent of
$\theta$, a non-zero value of $\theta$ in the above interval
must be chosen in order that $P5\pm$ exist (this pair of fixed points
vanish identically when $\theta = 0$). It is also the 
{\it only} IR-stable
fixed point in the three--dimensional coupling space 
spanned by the couplings ($V$,
$U_{0}$, $U_{\theta}$). This implies the following behavior: 
any point initially in the
basin of attraction of $P5+$ inevitably ends up at this fixed point as
larger and larger scales are probed: {\it no fine tuning is required}.

\section{Discussion and Conclusions.}

In this paper we have presented an {\em analytical} calculation of the
density-density correlation function for non-relativistic matter in a
FRW background cosmology. The calculation hinges on blending two
essential theoretical frameworks as input. First, we take and then
reduce the complete set of non-relativistic hydrodynamic equations
(Euler, Continuity and Poisson) for a self-gravitating fluid to a
single {\it stochastic} equation for the peculiar-velocity potential
$\psi$. The noise term $\eta(\vec x,t)$ models in a phenomenological but
powerful way different stochastic processes on various length and time
scales. Second, we apply the well-established techniques of the dynamical
Renormalization Group to calculate the long--time, long--distance
behavior of the correlation function of the velocity potential (which
relates directly, as we have seen, and under the conditions we have
specified,  
to the density-density correlation) as a
function of the stochastic noise source in the cosmological KPZ
equation.  

When we carry out a detailed RG fixed point analysis for our KPZ
equation, we find that simple white noise alone is not sufficient to
account for the scaling exponent $(1.6 \stackrel{<}{\sim} \gamma
\stackrel{<}{\sim} 1.8)$ inferred from observations of the galaxy-galaxy correlation
function. However, we can get close to this range for the observed exponent if the
cosmic hydrodynamics is driven by correlated noise. In fact, we need
only adjust the degree of spatial correlation of the noise to achieve this,
since the calculated value of the exponent $\gamma = 4 - 2\chi(\rho)$
associated with $P5+$ is independent of the degree of temporal
correlations, as encoded in $\theta$. Moreover, $P5+$ is the only
fixed point with this property, and is
simultaneously IR-attractive, both desirable properties from the
phenomenological point of view. This last property is very important,
since it means that the self--similar behavior of correlations is a
{\it generic} outcome of the dynamical evolution, rather than an atypical
property that the system exhibits only under very special conditions.

Thus we have learned 
that colored noise seems to play an important r\^ole
in the statistics of large scale structure. Now, {\it why} it is that
a non-zero $\theta$ and a $\rho \in (2.6, 2.75)$ are the relevant
noise exponents is a question one still has to ask. To answer it, one
would in principle have to derive the noise source itself starting
from the relevant physics responsible for generating the fluctuations. Here,
we content ourselves with the phenomenological approach.
Nevertheless, at some future point, it may be possible to say
more about the spectrum of noise fluctuations.

It must be pointed out that the Renormalization Group allows us to compute
only {\em asymptotic} correlations, i.e., after sufficient time
has elapsed so that the effect of noise completely washes out any
traces of the initial conditions in the velocity field or initial
density perturbations. Hence, there 
will be a time--dependent {\it maximum}
length scale, above which noise has not become dominant yet and
correlations therefore remember the initial conditions, as well as a
transition region between these two regimes. The length scale at which
this transition occurs depends on noise intensity and on the amplitude
of initial perturbations. Therefore, a physical model of the
origin of noise would enable us to predict this scale, but such a task
is well beyond the scope of the present paper.

It is also interesting that the Renormalization Group allows us
to compute the proportionality coefficient of the matter density
correlations, i.e., the quantity $r_{0}$ in
$\xi(r)=(r/r_{0})^{-\gamma}$, by means of the so-called {\it improved}
perturbation expansion (Weinberg \cite{Weinberg2}). This length scale $r_{0}$
marks the transition from inhomogeneity ($r \ll r_{0}$, $\xi(r) \gg
1$) to homogeneity ($r \gg r_{0}$, $\xi(r) \ll 1$). But $r_{0}$ is
also dependent on the noise intensity and other free parameters of our
model, so that a prediction is impossible without a deeper knowledge
of the noise sources. At this point, we should mention the recent
debate about the observational value of $r_{0}$: on the one hand there
is the viewpoint that homogeneity has been reached well within the
maximum scale reached by observations (Davis \cite{Davis}) ($r_{0}
\stackrel{<}{\sim} 50 Mpc$); on the other hand there is an opposing
viewpoint as expressed by
Labini et al. \cite{Pietronero} who argue that 
$r_{0} \stackrel{>}{\sim} 200 Mpc$. 
Either viewpoint can be easily incorporated into our model, by suitably
choosing the noise intensity. But if the latter value of 
$r_{0}$ turns out to be
correct, this would indicate within our model that the noise intensity
is really large and therefore, that noise and stochastic
processes played a much more important r\^ole in the History of the
Universe than has been thought to date.

\begin{acknowledgements}
The authors wish to thank  Arjun Berera, Fang Li-Zhi, Jim Fry,
Salman Habib, Herb Schnopper and Lee Smolin, for discussion.
\end{acknowledgements}


\end {document}